# Termination shock thermal processes as a possible source for the CMB low-order multipole anomalies: updated with observations

H.N. Sharpe

Bognor, Ontario, Canada
sh3149@brucetelecom.com

## Abstract

We discuss the possibility that the observed low-order multipole features of the cosmic microwave background radiation (CMB) all originate in the termination shock (TS) region of the heliosheath that surrounds the solar system. If the intrinsic CMB spectrum is assumed to be a pure monopole (2.73K) then thermodynamic processes occurring within the plasma region of the TS could imprint the observed power spectrum of the low-order multipoles and their alignment (the so-called "axis of evil") onto this background isotropic CMB. Conditions are outlined for the geometric shape of the TS region. A key requirement of this model is that the TS plasma be characterized as an optically thin graybody with non-LTE perturbations. Data from the ongoing Voyager missions is critical to this study. We present four significant recent observations in support of this ansatz.

## Introduction

In an earlier article we discussed the possibility that the observed quadrupole moment in the CMB power spectrum may in fact originate from the geometric distortion of the background CMB in the termination shock (TS) of the heliosheath [1]. In the present article we extend this heuristic model to a discussion of several anomalies reported in the low-order multipole power spectrum of the CMB. These include the apparent alignment of the quadrupole and the octupole ($l$=2,3) moments (and possibly $l$= 4 and 5) with the direction of the solar system's motion ("axis of evil" (AOE) [2,3,6])and the ecliptic plane as well as the low power in $l$=2,3 [2]. We also discuss the reported anomalies in non-blackbody departures of the CMB low order multipoles [4,5].

Termination shock physical processes can interact with the CMB radiation in one of two ways. The CMB photons can be absorbed and scattered by local radiative and kinetic processes as they propagate through the TS region (see [1] for a summary which also includes possible refraction at the TS). Or they can pass unaffected through an essentially transparent heliosheath. In this article we focus on the latter case. We assume that local thermal processes in the TS emit radiation that is simply superimposed on the background CMB. Since this radiation is emitted on the TS surface it should acquire the geometric properties of the TS. Accordingly, the observed CMB spectrum inside the solar system will reflect the multipole distortions of the TS. We outline the physical and geometric conditions which must be satisfied at the TS to explain the observed CMB low-order multipole anomalies with a local rather than cosmic model.

## Model

We start with the general transfer equation for a pencil of radiation propagating through an optically thin region characterized by optical depth $\tau_{\nu}$ and source function $S_{\nu}$ and assume isotropy:

$$I_{\nu}(\tau_{\nu}) = I_{\nu}(0) + \tau_{\nu} S_{\nu} \qquad \tau_{\nu} << 1 \qquad (1)$$

We identify $I_{\nu}(0)$ with the CMB background brightness spectrum and $\tau_{\nu} S_{\nu}$ with the additional brightness from the TS region.

## Geometric Structure

First we address the CMB blackbody low-order multipoles. If we assume the CMB is perfectly isotropic with only a monopole $T_0$=2.73K, then all the low order multipoles must arise from geometric distortions of the TS surface relative to a spherical surface. In [1] we showed that a slight distortion to a prolate ellipsoid of revolution could account for the quadrupole rms amplitude of ~ $14\mu$ K. This geometric distortion was shown to be consistent with preliminary observations of an asymmetric TS from the Voyager spacecraft [7]. However, an ellipsoid of revolution does not contain odd-numbered harmonics. Therefore a more generalized ellipsoidal distortion to the TS must be considered:

$$\frac{x^2}{a^2} + \frac{y^2}{b^2} + \frac{z^2}{c^2} = 1 \qquad (2)$$

where a,b,c, are chosen to satisfy the multipole constraints on $T_l/T_0$, l=2,3,4 and possibly $l$=5. This is an optimization problem. The result will be an ellipsoidal surface which satisfies the magnitudes of multipoles 2 to 5 and their alignment with the Sun's motion and the ecliptic plane since the principal distortion of the TS is in the Sun's direction through the ISM. It should also be noted that this ellipsoid will generate a dipole moment (l=1). This geometric moment should however be subsumed in the much larger Doppler dipole. With sufficient computing power a more generalized and rigorous treatment of this problem would involve a full spherical harmonic expansion of the ellipsoidal surface with a constrained optimization of the $C_{nm}$ for the observed multipoles l=2 to 5.

## Physical Mechanisms

For the geometric distortions of the TS to manifest as blackbody multipoles against an otherwise isotropic CMB background, the physical mechanism operating in the TS region must be in local thermodynamic equilibrium (LTE). First we demonstrate how this process could work and then discuss its relevance.

In equation (1) we make the following key assumptions:

1. The source function is Planckian (LTE): $S_{\nu} = B_{\nu}$ EMBED Equation.3

2. The absorption coefficient (opacity) may be represented as a graybody (independent of frequency) with a small departure from grayness, $\delta_{\nu}(\theta)$ that in general could depend on direction:

$$\tau_{\nu} \rightarrow \bar{\tau}(1 + \delta_{\nu}(\theta)) \qquad \bar{\tau} \equiv \bar{\alpha} L \;<< 1 \text{ and } \delta_{\nu} << 1$$

Assumption 2 preserves the blackbody profile of the emitted thermal radiation while allowing for small deviations. With these assumptions and using Rayleigh-Jeans equation (1) becomes:

$$T_{\nu}^{obs} = T_{cmb} + \bar{\tau}T^{TS} + \bar{\tau}T^{TS}\delta_{\nu}(\theta) \tag{3}$$

The first term on the rhs is the CMB monopole 2.73K. The second term is the graybody Planckian perturbation to the CMB monopole caused by the TS geometric distortion. It may be thought of as a normalization constant. $T^{TS}$ is the effective local kinetic temperature in the TS.

The third term on the rhs is the small departure from a blackbody spectrum due to non-grayness of the thermal radiation. If we assume it has no directional dependence on the TS surface then this non-blackbody contribution will inherit the same geometric distortion as the blackbody perturbations. This could be a cause for the observed alignment of the non-blackbody multipoles with the overall CMB multipoles [4,5].

The assumption of graybody thermal radiation is key to this model. It requires that detailed balance holds for collisional processes in the TS region, which in turn requires that the particle distribution function is Maxwellian [8]. Under these conditions a local kinetic temperature can be defined and LTE is valid. Radiative processes can compete with collisions but these are not expected to be a factor in the TS.

Since the TS is a shock interface between the supersonic solar wind and the heliosheath region, the shock can disturb the tendency towards LTE. Several recent reports discuss the findings of the Voyager spacecraft which have now penetrated the TS region [9,10,11]. They show a chaotic, turbulent magnetized plasma far from LTE. Nevertheless, it may still be possible to model this region globally as a perturbed Maxwellian graybody distribution. On-going research into this difficult problem is continuing. *If an appropriate distribution function for the TS can be developed it will provide an important local alternative explanation to cosmic models for the observed low-order CMB anomalies*. In addition the TS model can continue to be refined with more data as the Voyager spacecraft penetrate deeper into the heliosheath.

## Diego et al "Toy" Model

In the model just presented we assumed an isotropic monopole for the CMB background radiation. All observed blackbody and non-blackbody low-order deformations were attributed to geometric distortions of the TS and in situ LTE processes ( with small departures from Planckian). In [4] Diego et al assume a cosmological origin of the full CMB multipole spectrum, but attempt to explain the low quadrupole power ( low relative to inflation model predictions) and its alignment with the octupole and the ecliptic plane in terms of potential sources of contamination. They present a quasi-blackbody "toy" model normalized such that its non-blackbody amplitude is approximately $10\mu$ K in the V+W-2Q band (after filtering with a 7 deg Gaussian). They then demonstrate that this "toy" model possesses a quadrupole which is anti-correlated with the WMAP5 quadrupole. When subtracted from the observed quadrupole the result better approximates the value expected from theory. The octupole is not affected by their model since significant power is only found in the even moments. The low order multipoles also seem to come into better alignment.

While this "toy" model is capable of explaining the observed anomaly, Diego et al acknowledge that a physical justification for its existence is lacking. We conclude this note with a possible justification for their "toy" model.

We found earlier that it was necessary to generalize the TS distortion from a simple ellipsoid of revolution in order to potentially explain all the low-order multipoles of the CMB and their alignments. However, if we are only interested in the even multipoles, specifically the quadrupole, then we may just adopt the procedure discussed in [1] and return to an ellipsoid of revolution for the TS distortion. Then we may write equation (1) as:

$$I_\nu^{obs} = I_\nu(0) + B_\nu[\bar{\tau}(1+\delta_\nu)] \tag{4}$$

where $I_\nu(0)$ is the full CMB multipole background and the second term represents the quasi-blackbody distortion due to non-LTE processes on the TS as discussed above. The normalization can be chosen such that this term satisfies the same $10\mu$ K constraint in the V+W-2Q band. Since the distortion originates on the TS it will share its geometric properties of only possessing even multipoles. Hence equation (4) could be a physical justification of the "toy" model proposed by Diego et al.

## Recent Supportive Observations

1. A key requirement of my model is that the heliosphere possess a closed ellipsoidal-like geometry rather than a magnetosphere-like open heliotail structure. Now a recent paper using the Voyager and Cassini data has concluded that a diamagnetic bubble-like heliosphere is the appropriate model with few tail-like features [13]. Low-order departures of this spherical bubble-like shape could imprint similar distortions on an assumed CMB monopole. It would also be interesting to fit a full spherical harmonic distribution to a perfect sphere such that this distribution exactly matches the FULL CMB harmonic power spectrum and not just the low-order multipoles.

2. The two Voyager spacecraft, which have now entered the ISM, both show the heliosheath plasma to be much hotter and denser than previously expected [14]. If these observations can be extrapolated around the entire heliosheath they could support the model requirement for an optically thin graybody.

3. The observed non-Doppler dipole anomaly in the CMB has been interpreted as possibly intrinsic to the CMB itself. That is, one side of the sky may actually be hotter than the opposite side [15]. However a more natural explanation is to suggest that the compressed heliosheath at the bow is hotter (more dense) relative to the heliosheath plasma in the tail. That is, the observed non-Doppler dipole anomaly across the sky is imprinted on the CMB background radiation by the motion of the closed ellipsoidal heliosheath through the ISM as observed from the Earth.

4. A recent paper by de Graff et al claims that the missing baryon mass can be explained by the tendrils of the warm-hot intergalactic medium (WHIM) that connect across inter-galactic space [16]. They use the CMB Planck data to show that this hot, ionized gas can interact with the CMB photons to distort or refract the CMB spectrum. These inter-galactic plasma filaments are much thinner and less dense (reduced optical depths) than the heliosheath plasma. It would thus seem that a similar refraction of the CMB should be expected across the heliosheath thereby imprinting the low-order harmonics of the heliosheath geometry on the observed CMB spectrum.

Taken together, these diverse observations suggest that the ansatz presented in this paper could be a viable mechanism for imprinting at least the low-order multipoles on an otherwise uniform background CMB.

## Conclusions

The Voyager spacecraft have provided a unique opportunity to characterize the dynamical and radiative processes in the heliosheath which surrounds the solar system. Since the CMB is viewed through this "window" it should be expected that these processes may leave their imprint on the CMB spectrum. A rigorous thermodynamical model is required for the heliosheath plasma and the termination shock region. If a dusty plasma model can be developed which characterizes this region as an optically thin graybody with non-Maxwellian perturbation distributions, then most of the low-order multipole features of the CMB spectrum could be explained by a local rather than cosmic origin. The high order multipole components should be studied with a full MHD turbulence model for the TS region to determine if the observed acoustic peaks can also be given a local interpretation. The observed CMB polarization could also have its origin in heliosheath radiative processes like Thomson and Compton scattering and inverse Compton scattering off suprathermal ions/electrons. Synchrotron radiation could also contribute to this polarization as well as providing a possible explanation for other observed anomalies [12]. Recent observational data from several diverse sources support the proposed ansatz as an alternative to a cosmological origin of the CMB multipole spectrum.

al Oct.5 2017. arXiv:1709.10378v2.